\documentclass[conference, a4paper]{IEEEtran}
\ifCLASSINFOpdf
 \else
  \fi
\usepackage{amssymb}
\usepackage{amsfonts}
\usepackage{amsmath}
\usepackage{fancyhdr}
\usepackage{epsfig}
\usepackage{cite}
\usepackage{graphicx}
\usepackage {longtable}
\usepackage[linesnumbered,vlined,ruled]{algorithm2e}
\newcommand{\Cc}{\ensuremath{\mathcal  C}}

\newcommand{\Gc}{\ensuremath{\mathcal  G}}
\newcommand{\Hc}{\ensuremath{\mathcal  H}}

\newcommand{\Pc}{\ensuremath{\mathcal  P}}
\newcommand{\Sc}{\ensuremath{\mathcal  S}}

\newcommand{\dr}{\ensuremath{\mathrm{d}}}

\newcommand{\bF}{\ensuremath{\mathbb{F}}}

\newtheorem{proposition}{Proposition}[section]
\newtheorem{lemma}{{ Lemma}}[section]

\newcommand{\bit}{\begin{itemize}}
\newcommand{\eit}{\end{itemize}}
\newcommand{\bcor}{\begin{cor}}
\newcommand{\ecor}{\end{cor}}
\newcommand{\beq}{\begin{equation}}
\newcommand{\eeq}{\end{equation}}
\newcommand{\beqn}{\begin{equation*}}
\newcommand{\eeqn}{\end{equation*}}
\newcommand{\bea}{\begin{eqnarray}}
\newcommand{\eea}{\end{eqnarray}}
\newcommand{\bean}{\begin{eqnarray*}}
\newcommand{\eean}{\end{eqnarray*}}
\newcommand{\ben}{\begin{enumerate}}
\newcommand{\een}{\end{enumerate}}
\newcommand{\bdefn}{\begin{definition}}
\newcommand{\edefn}{\end{definition}}
\newcommand{\bnote}{\begin{note}}
\newcommand{\enote}{\end{note}}
\newcommand{\bprop}{\begin{proposition}}
\newcommand{\eprop}{\end{proposition}}
\newcommand{\blem}{\begin{lemma}}
\newcommand{\elem}{\end{lemma}}
\newcommand{\bthm}{\begin{theorem}}
\newcommand{\ethm}{\end{theorem}}
\newcommand{\dc}{doubling construction}

\newcommand{\ag}{\ensuremath{PG(4,2)}}

\newcounter{arno}
\newcounter{rno}
\newtheorem{definition}{Definition}[section]
\newtheorem{theorem}{ Theorem}[section]

\begin{document}
\title{Optimal Binary $(5,3)$ Projective Space Codes from Maximal Partial Spreads}

\author{\IEEEauthorblockN{Anirban Ghatak}
\IEEEauthorblockA{Department of Electrical Communication Engineering\\
Indian Institute of Science\\
Bangalore, Karnataka, India 560012\\
Email: aghatak@ece.iisc.ernet.in}}

%\IEEEoverridecommandlockouts
%    \IEEEpubid{\makebox[\columnwidth]{978-1-5090-2361-5/16/\$31.00~\copyright~2016~IEEE \hfill}
%    \hspace{\columnsep}\makebox[\columnwidth]{ }}

\maketitle

\begin{abstract}
Recently a construction of optimal non-constant dimension subspace codes, also termed projective space codes, has been reported in a paper of Honold-Kiermaier-Kurz. Restricted to binary codes in a $5$-dimensional ambient space with minimum subspace distance $3$, these optimal codes may be interpreted in terms of maximal partial spreads of $2$-dimensional subspaces. In a parallel development, an optimal binary $(5,3)$ code was obtained by a minimal change strategy on a nearly optimal example of Etzion and Vardy. In this paper, we report several examples of optimal binary $(5,3)$ codes obtained by the application of this strategy combined with changes to the spread structure of existing codes. We also establish that all our examples lie outside the framework of the construction of Honold \emph{et al.}
\end{abstract}
\section{Introduction}
The notion of subspace coding for errors and erasures in random networks was introduced in the papers of K\"{o}tter \emph{et al.} (\cite{KK, SKK}). There exists a rich body of literature dealing with the encoding and decoding of constant dimension codes, (for instance, references \cite{KoKu} -\cite{GT}) where all the codeword subspaces have the same dimension. An extensive list of open problems in this field and relevant references are to be found in Etzion's survey \cite{Etzion}. In comparison, the more general problem of the construction of non-constant dimension subspace codes, termed projective space codes, still has a lot of scope. \\
Etzion and Vardy \cite{EV} proved a Gilbert-Varshamov-type lower bound and a linear programming upper bound - assuming a given \emph{dimension distribution} - on the size of projective space codes. In addition, they gave an example of a nearly optimal binary code of minimum \emph{subspace distance} $ 3 $ in a $ 5 $-dimensional ambient space, where the subspace distance between two subspaces $ U, V $ of an ambient projective space is defined in \cite{KK} as:\\
 $ {\dr}_s(U,V) :=\dim U+ \dim V - 2\dim (U \cap V) $\\
In \cite{KhK} and \cite{AG}, the construction of constant dimension Ferrers-diagram rank-metric codes in \cite{ES} has been adapted for constructing projective space codes. Another approach ( for instance, in \cite{ES, ST, HKK}) involves \emph{puncturing} known constant dimension codes and adding suitable subspaces to increase the overall code-size.\\
Honold, Kiermaier and Kurz \cite{HKK} have described a method for generating optimal $(5,3)$ codes over $\bF_q$, termed the \emph{point-hyperplane shortening} of a constant dimension lifted Gabidulin code \cite{Gab}, augmented by additional subspaces maintaining the minimum subspace distance. Over $\bF_2$, their construction, henceforth referred to as the HKK Construction, results in optimal binary codes of size $18$. The authors have shown that the basic building blocks of these codes are \emph{maximal partial spreads} of $2$-dimensional subspaces in a $5$-dimensional ambient space.\\
An optimal $(5,3)_2$ projective space code was reported in \cite{AG1}, which was obtained by a strategy of minimal changes to a nearly optimal code given by Etzion-Vardy in \cite{EV}. Both the Etzion-Vardy (EV) code and the optimal variant can be described in terms of maximal partial spreads of $2$-dimensional subspaces. As a natural extension, we have searched for optimal and nearly optimal codes composed of partial spreads which are derived from those used in these two codes. In doing so, we have used results from the classification of such maximal partial spreads as presented in \cite{RS} (cf. \cite{HKK14, HKK}).\\
The main contribution of this paper consists of two pairs of optimal binary $(5,3)$ codes obtained by applying the minimal change strategy on some nearly optimal codes. These nearly optimal codes were, in turn, constructed from the EV code and its variant by replacing the constituent maximal partial spreads with their \emph{opposite} spreads. In addition, we report a third pair of optimal codes obtained by a combination of some of the codes mentioned above. Finally we establish that all the codes reported in this paper fall outside the framework of the HKK construction.\\
The organization of this paper is as follows. The next section provides some relevant results regarding optimal binary $(5,3)$ projective codes and, in that context, maximal partial spreads of $2$-dimensional subspaces in an ambient $5$-dimensional space. The third section first reviews the minimal change strategy which produces the optimal variant from the EV code and then proceeds to describe the two pairs of opposite regulus optimal codes. In the fourth section we examine the HKK construction and provide a comparison with our codes. We conclude with a discussion on several unanswered questions related to the results presented.\\
\emph{Notation:} We use $\bF_q$ to denote the finite field of $q$ elements and $PG(n-1,q)$ to denote the $(n-1)$-dimensional projective geometry over $\bF_q$. The latter is equivalent to the set of all subspaces of the canonical $n$-dimensional vector space $\bF_{q}^{n}$ ordered by the incidence relation $\subseteq$ (\cite{HKK}). Hence we interchangeably refer to the $2$-subspaces of $\bF_{2}^{5}$ as \emph{lines} of $PG(4,2)$, $3$-subspaces as \emph{planes}. A hyperplane of $\bF_{q}^{n}$ is identified with $PG(n-2,q)$. \\
As used in \cite{RS}, we adopt the following compact representation for vectors in $\bF_{2}^{5}$. Identify the canonical basis vectors $e_i$ with the indices $i = 1,2,\cdots, 5 $. An arbitrary vector is denoted by the tuple of the constituent basis vectors; for instance, the vector $10101$ is denoted by the tuple $135$. The vector $11111$ is denoted $u$ - the tuple $4u$ thus stands for the vector $11101$.
\section{Optimal $(5,3)$ Codes and Maximal Partial Spreads}\label{sec:bg}
In this section we discuss the properties of optimal $(5,3)$ codes and examine how they can be realized using maximal partial spreads of $2$-dimensional subspaces in a $5$-dimensional vector space. In \cite{RS}, a complete classification of such maximal partial spreads, i.e. maximal partial line spreads in $PG(4,2)$, is presented. We briefly review the relevant results and concepts which will be used in subsequent sections. 
\subsection{Optimal $(5,3)$ Codes and the HKK Construction }
A $(5,3)_q$ subspace code is a subset of $PG(4,q)$ which can include subspaces of all possible dimensions with pairwise minimum subspace distance $3$. In \cite{HKK}, all possible dimension distributions as well as the achievable size of optimal $(5,3)_q$ subspace codes are specified. The size of such a code is $2q^3 + 2$ and a possible realization is the union of $q^3 + 1$ lines ($2$-subspaces) and an equal number of planes ($3$-subspaces). Other realizations involve replacing one line by a $1$-dimensional point and/or replacing one plane by a hyperplane ($4$-subspace) in the above. If we set $q=2$ in the first case, we have $9$ lines and an equal number of planes.\\
\emph{HKK Construction:} The $(6, 3, 2)$ lifted Gabidulin code $\Gc$ is a set of $3$-subspaces of a $6$-dimensional vector space ($PG(5,q)$), obtained by the K\"{o}tter-Kschischang lifting \cite{KK} of a Gabidulin code of $3 \times 3$ matrices with rank distance ${\delta}_r = 2$. The minimum subspace distance of $\Gc$ is ${\dr}_s = 2{\delta}_r = 4$. The \emph{point-hyperplane shortening} is performed on the set of subspaces of $\Gc$, plus two additional $3$-subspaces (planes of $PG(5,q)$) at subspace distance $4$. The shortening point is outside the special plane $\Sc$ of $PG(5,q)$ disjoint from all codewords of $\Gc$ (cf. Section $2.4$, \cite{HKK}) and the shortening hyperplane contains neither the special plane nor the point. The simultaneous shortening of $\Gc$ with respect to such a point $P$ and a hyperplane $H = PG(4,q)$ is described as:
\beqn
\Gc \lvert^{P}_{H} = \{ X \in \Cc \, \lvert \, X \subset H \}\cup \{Y\cap H, Y \in \Cc \, \lvert \, P \in Y \}
\eeqn 
The lifted code provides sets of $2$- and $3$-dimensional subspaces, each of size $q^3$, via shortening; one of the additional pair of $3$-subspaces furnishes a $2$-subspace via shortening, the other is included as a codeword. Thus we have a $(5, 3)_q$ code of size $2q^3 + 2$ with dimension distribution $(2,3)$.
\subsection{Maximal Line Spreads in $PG(4,2)$}\label{subs:sprd}
A $spread$ in a vector space is a collection of non-intersecting subspaces which completely partitions the ambient space; a spread is $partial$ when it does not cover the entire ambient space. In \cite{RS} a complete classification of maximal partial spreads in $PG(4,2)$ is given in terms of the \emph{regulus patterns} of the spreads. Following \cite{HKK14}, we have the definition:
\bdefn
 A \emph{regulus} in $PG(4,2)$ is a set of three pairwise skew lines spanning a solid ( hyperplane ).
\edefn
The number of reguli contained in maximal partial spreads $S_r$ of size $r$ is given in Theorem $2.4$ of \cite{RS} as follows.
\bthm
If $S_r$ is a maximal partial spread of $r$ lines in $PG(4,2)$ then one of the following holds:\\
(i) $r = 5$, $N_5 = 10$;  (ii) $r = 7$, $N_7 = 4$; (iii) $r = 9$, $N_9 = 4$.\\
Moreover each of the above possibilities is realized. $\blacksquare$  
\ethm 
It follows that in $PG(4,2)$, maximal partial line spreads can have sizes $5,7, 9$  - hence we term a partial spread of size $9$ as a \emph{maximum} partial spread (Mps) (cf. \cite{DF}). An Mps of lines in $PG(4,2)$ has $4$ reguli, i.e. there are $4$ sets of three lines $\{ l_1 , l_2 , l_3 \}$, each set spanning a hyperplane. The intersections among these $4$ sets determine the \emph{regulus pattern} of the Mps. Any Mps of lines in $PG(4,2)$ exhibits one of the following three types of regulus patterns.
\ben
\item Type X : All $4$ reguli share a common line.
\item Type I$\Delta$ : Three of the reguli form a ``triangle", i.e. share two distinct lines with two other reguli; one regulus has no intersection with the other three.
\item Type E : Three reguli share a distinct line each with the fourth regulus.
\een
\emph{New Spreads from Opposite Reguli:}\\
A regulus in $PG(4,2)$ can equivalently be thought of as a set of $3$ lines covered pointwise by a second set of $3$ mutually skew lines (\cite{BJJ}). The second set is called the \emph{opposite regulus}, and new spreads can be obtained from existing ones by replacing a particular regulus with its opposite regulus \cite{RS}.\\
$\bullet$ \emph{Between I$\Delta$ Spreads:} An $S_9$ of type I$\Delta$ is constructed from a size-$6$ spread of type $\Delta$, say $S_6(\Delta)$, as follows. As indicated, $S_6(\Delta)$ already has three reguli forming the $\Delta$. The complementary set of such an $S_6$ in $PG(4,2)$ can be expressed as the union of a hyperbolic quadric $\Hc$ and a $4$-set. The $4$th regulus of this $S_9$ can be chosen from the two opposite reguli which cover the $9$ points of $\Hc$. So replacing the special regulus, forming the `I', with its opposite regulus changes an I$\Delta$ spread to its opposite spread, again of type I$\Delta$. \\ 
$\bullet$ \emph{Between Spreads of Type X and E:} A type X Mps $S_9$ can be generated from a \emph{cyclic} $S_6$ in a manner similar to the generation of a type I$\Delta$ $S_9$ from an $S_6(\Delta)$ (\cite{RS}). A cyclic $S_6$ is a set of $6$ mutually skew lines which are cyclically fixed by an element of $GL(5,2)$ of order $6$. The complement of this $S_6$ in $PG(4,2)$ is the disjoint union of a plane $\alpha $ and two mutually skew lines $l_7 $ and $l_8 $. Any line $l_9 \subset \alpha$ completes an $S_9$ of type X along with $l_7, l_8$. The selection of these lines is dictated by the chosen element of $GL(5,2)$: $l_9$ is fixed and the pair $(l_7,l_8)$ transposed by the element. If the regulus $\{l_7, l_8, l_9 \}$ is replaced with its opposite regulus, we obtain a type E spread from the original type X spread.\\ 
Going from a type E spread to the opposite type X spread entails the replacement of the shared regulus of the type E spread by its opposite regulus.
%\beg Consider the spread $S1$, the Mps in the EV code of type X. Evidently we have to find a permutation of the basis set $\{1,2,3,4,5\}$ which fixes $l_9 = \{134,2,5u \}$, cyclically permutes six of the remaining $8$ lines, and transposes the last pair. The following permutation, described as: $(12345)\rightarrow (3,5u,24,34,5)$, acts on the lines of $S1$ as: $(l_6 \,l_1\,  l_4\, l_2\, l_5\, l_3)\,(l_7\, l_8)\,(l_9)$.\\
%Hence the lines $l_1 = \{35,45,34 \} $, $l_2 = \{13,235,125 \}$, $l_3 = \{14,4u,1u \}$, $l_4 = \{245,345,23 \}$, $l_5 = \{135,3u,234 \}$, $l_6 = \{15,4,145 \}$ form the cyclic $S_6$, while $l_7 = \{12,3,123 \}$ and $l_8 = \{24,1,124 \}$ are the two lines making up the fourth regulus with $l_9$.\\
%To construct the Mps of type E from $S1$, one replaces the lines of the fourth regulus $\{l_7, l_8, l_9\}$ with $l_{7}^{o} = \{12, 1, 2 \}$, $l_{8}^{o} = \{3, 124, 5u\}$ and $l_{9}^{o} = \{123, 24, 134\}$. The regulus structure of the resulting Mps, denoted $S1^{o}$, is : $r_1 = \{l_1, l_5, l_{7}^{o}\}$, $r_2 = \{l_2, l_6, l_{9}^{o}\}$, $r_3 = \{l_3, l_4, l_{8}^{o}\}$ and $r_4 = \{l_{7}^{o}, l_{8}^{o},l_{9}^{o}\}$. Evidently $S1^{o}$ is of type E, with each of the three lines of $r_4$ shared with exactly one of the other three reguli.
%\eeg
\subsection{The Doubling Construction}
An optimal $(5,3)_2$ code with dimension distribution $(2,3)$ can be expressed as the union: $S1\cup (S2)^{\perp}$, where $S1, S2$ are Mps's in $\ag$. That is to say, the set of $3$-subspaces (planes of $\ag$) can be viewed as the element-wise dual of another Mps of $2$-subspaces (lines of $\ag$). This follows from Lemma $13$ of \cite{EV}, stated below, restricted to constant dimension codes. Recall that in \cite{EV}, the projective space $ \Pc_{q} (n) $ is defined to be the collection of all possible subspaces of an $n$-dimensional vector space $V$ over $\bF_q$. Also, $\Cc \subseteq \Pc_{q} (n)$ is an $(n,M,d)$ code if $\lvert \Cc \rvert = M$ and the minimum subspace distance between any $U,V \in \Cc$ is $d$.   
\blem \label{lem:du}
If $\Cc$ is an $(n,M,d)$ code in $ \Pc_{q} (n) $, then its orthogonal complement ${\Cc}^{\perp}:= \{ C^{\perp} \, \lvert \, C \in \Cc  \} $ is also an $(n,M,d)$ code.
\elem
This construction of optimal codes as the union of a maximal spread and the complement of another, has been termed the \emph{doubling construction} by Honold \emph{et al.} \cite{HKK}. Moreover, if a pair of spreads $S1$ and $ S2$ yields an optimal $(5,3)_2$ code via the {\dc}, the `dual' union is another optimal code, as stated in the following Proposition \ref{prop:du}. This is a direct consequence of Lemma \ref{lem:du}. 
\bprop\label{prop:du}
If two maximum partial spreads $S1$ and $S2$ of $k$-subspaces in $PG(2k,q)$ form an optimal $(2k+1,2k-1)_q$ code via the {\dc} as $S1 \cup (S2)^{\perp}$, the dual union $ S2 \cup (S1)^{\perp}$ is an optimal code as well. 
\eprop
\section{The Optimal $(5,3)_2$ Projective Space Codes}
In this section we present several optimal $(5,3)_2$ codes. Our `method' may be summarized as follows. We start with an existing (nearly) optimal code which is the union of an Mps and the complement of another. Replacing the special reguli of the Mps's with the opposite reguli replaces three skew lines with their transversals. So we get new spreads which are very `close' to the originals, with the point set and two-thirds of the lines unchanged. If we obtain a pair of new Mps's with only a single conflict between them, we try to obtain an optimal code by the minimal change strategy.
\subsection{The Minimal Change Strategy: MEV Code and its Dual}
The EV code is a nearly optimal code of size $17$ which consists of two Mps's, denoted $S1, S2$, having regulus types X and I$\Delta$, respectively. The $2$-subspaces or lines in the respective spreads are denoted by $l_i(Sj),\, i = 1,2, \cdots, 9; \, j =1,2$, and are completely described as triples of vectors. The $3$-subspaces or planes are described later in terms of triples of linearly independent spanning vectors.\\
$S1 = \{35,45,34 \}, \{13,235,125 \}, \{14,4u,1u \},$\\$ \{245,345,23 \}, \{135,3u,234 \}, \{15,4,145 \}, \{12,3,123 \}, $\\$ \{24,1,124 \}, \{134,2,5u \}$.\\
$S2 = \{35,45,34 \}, \{14,4u,1u \}, \{13,235,125 \}, \{3, u, 3u\},$\\$ \{245,2u,123 \}, \{15,25,12 \},  \{134,2,5u \},\{23,4,234 \},$\\$ \{24,1,124 \}.$\\
To get an optimal code, we restrict our efforts to \emph{minimal} changes in the following sense. To resolve the distance discrepancy among two codewords $c(1), c(2)$, we change the row vectors of $c(1)$ or the row vectors of $c(2)$, choosing only one subspace at a time. Moreover, among the two chosen row vectors in the spanning matrices of each subspace, we change one particular row vector at a time. The replacements are chosen only among vectors which preserve the Schubert cell signature of each subspace. The details of constructing the MEV (modified EV) code have been given in \cite{AG1} and are briefly recounted here for completeness.\\
The line $l_2(S1) $ is a subspace of the orthogonal complement of the line $l_5(S2)$; hence it was attempted to modify either of them by changing a single vector at a time. The change that worked was the substitution of the tuple $001$ in the rightmost $3$ bits of the $2$-vector $235$ of $l_2(S1)$. This produced the modified line $\{13, 25, 4u \}$. The above substitution was suggested by the fact that the tuple $001$ was the only one not used in the rightmost $3$ bits of the $2$-vectors in any of the lines in $S1$. But this change results in non-trivial intersection between $l_2(S1)$ and $l_3(S1)$. However, the tuple $101$ can be substituted in the $2$-vector of $l_3(S1)$, forming the line $\{ 14, 235, u\}$. The modified spread $S1'$ with the above lines for $l_2(S1'), l_3(S1')$, and $l_i(S1')= l_i(S1)$ for all other $i$, forms an optimal code (modified EV or MEV code) with $S2$ via the {\dc}.
The regulus pattern for $S1'$ is given by: $r_1 = \{l_1, l_5, l_9\}$, $r_2 = \{l_2, l_4, l_6\} $, $r_3 = \{l_3, l_5, l_7\}$ and $r_4 = \{l_7, l_8, l_9\}$, where $l_i = l_i(S1')$ for all $i$. It is evident that $r_1, r_3 ,r_4$ form the $\Delta$, while $r_2$ is the regulus `I'.\\
\emph{The MEV Code and its Dual:}
\begin{itemize}
\item MEV code:\\
$S1'$: $ \{35,45,34 \}, \{13,25,4u \}, \{14,235,u \}, $ \\$\{245,345,23 \} \{135,3u,234 \}, \{15,4,145 \}, \{12,3,123 \},$\\$ \{24,1,124 \}, \{134,2,5u \}$.\\
$(S2)^{\perp}$: $\langle 1,2,345 \rangle, \langle 145,25,35 \rangle,\langle 135,25,4 \rangle, \langle 15,25,45 \rangle,$\\$\langle 13,235,45 \rangle, \langle 125,3,4 \rangle,\langle 14,34,5 \rangle, \langle 1,23,5 \rangle,\langle 24,3,5 \rangle $.
\item The dual code:\\
 $S2$: Described earlier.\\
$(S1')^{\perp}$: $\langle 1,2,345 \rangle, \langle 13,25,4 \rangle,\langle 14,25,35 \rangle, \langle 1,235,45 \rangle,$\\$\langle 15,24,345 \rangle,\langle 15,2,3 \rangle, \langle 12,4,5 \rangle,\langle 24,3,5 \rangle, \langle 14,34,5 \rangle $.
\end{itemize}
\subsection{Opposite Regulus Codes}
In this subsection we present two pairs of optimal codes obtained by the application of the minimal change strategy on nearly optimal codes that result from replacing spreads of existing codes by their opposite spreads.
\subsubsection{Opposite Regulus Code from the MEV Code} 
Replacing the distinguished reguli, i.e. the `I' in the I$\Delta$ pattern of the spreads $S2$ and $S1'$ in the MEV code, with the opposite reguli, we obtain two new Mps's $S2^{o}$ and $S1'^{o}$ as follows.\\
$\bullet$ The reguli of $S2$ are given by: $r_1 = \{l_1, l_6, l_2\}$, $r_2 = \{l_2, l_4, l_3\} $, $r_3 = \{l_3, l_5, l_1\}$ and $r_4 = \{l_7, l_8, l_9\}$. The lines $l_1, \cdots, l_6$ form the $S_6(\Delta)$, while $r_4$ is the distinguished regulus which provides one set of generators of the quadric $\Hc$. Describing the quadric as an array of points, we have:
\beqn
\Hc = \begin{pmatrix}
2  & 5u  & 134 \\
4  & 234  & 23 \\
24  & 1  & 124 
\end{pmatrix}      
\eeqn
The lines of $r_4$ form the rows of $\Hc$, and the columns constitute the lines of the opposite regulus ${r_4}^{op}$. Therefore a second spread of type I$\Delta$, denoted $S2^{o}$, is formed by replacing the lines $l_7,l_8,l_9$ of $S2$ with:
$l_{7}^{o} = \{ 2,4, 24 \}$, $l_{8}^{o} = \{ 5u, 234, 1 \}$, and $l_{9}^{o} = \{ 134,23, 124 \}$.\\ 
$\bullet$ The quadric array formed by the distinguished regulus of $S1'$ is
\beqn
\Hc = \begin{pmatrix}
13  & 25  & 4u \\
345  & 245  & 23 \\
145  & 4  & 15 
\end{pmatrix}      
\eeqn  
Thus $S1'^{o}$ is obtained from $S1'$ by replacing the lines of the regulus $\{l_2, l_4, l_6\}$ with $l_{2}^{o} = \{13, 345, 145 \}$, $l_{4}^{o} = \{25, 245, 4\}$ and $l_{6}^{o} = \{4u, 23, 15\}$.\\ 
The code $ S1'^{o}\cup (S2^{o})^{\perp}$ is sub-optimal, with a single conflict between one $2$-subspace and one $3$-subspace. Specifically, $l_{4}^{o}:= l_4(S1'^{o})$ is contained in the dual subspace of $l_3(S2^{o})$. Attempting the $3$-tuple replacement in the $2$-vector as in the previous case, we find that the possible replacements are $111$ and $011$. Using $111$ in $l_{4}^{o}$ yields a modified $l_{4}^{o} = \{1u, 235, 4\}$ which is no longer non-intersecting with $l_3(S1'^{o}) = \{14, 235, u\}$. Replacing the tuple in the $2$-vector with $011$ yields a modified $l_3(S1'^{o}) = \{14, 245, 125 \}$. With these modifications, the pair of spreads $mS1'^{o}$  (modified $S1'^{o}$) and $S2^{o}$, of types X and I$\Delta$, form another optimal code OR1: $mS1'^{o} \cup (S2^{o})^{\perp}$. The dual union $ S2^{o} \cup (mS1'^{o})^{\perp}$ is another optimal code.\\
\emph{The Code OR1 and its Dual}
\begin{itemize}
\item $mS1'^{o}$: $ \{35,45,34 \}, \{13,345,145 \}, \{14,245,125 \},$\\$ \{235,4,1u \}, \{135,3u,234 \}, \{4u,23,15 \}, \{12,3,123 \},$\\$ \{24,1,124 \}, \{134,2,5u \}$.\\
$(S2^{o})^{\perp}$: $\langle 1,2,345 \rangle, \langle 145,25,35 \rangle,\langle 135,25,4 \rangle, $ \\ $\langle 15,25,45 \rangle,\langle 13,235,45 \rangle, \langle 125,3,4 \rangle, \langle 1,3,5 \rangle, $ \\ $\langle 24,34,5 \rangle,\langle 14,234,5 \rangle $.
\item The dual code:\\
$S2^{o}$: $\{35,45,34 \}, \{14,4u,1u \}, \{13,235,125 \}, $ \\$ \{3, u, 3u\}, \{245,2u,123 \}, \{15,25,12 \},  \{2,4,24 \},$\\$ \{1,234,5u \}, \{134,23,124 \}.$ \\
$(mS1'^{o})^{\perp}$: $\langle 1,2,345 \rangle, \langle 135,2,45 \rangle,\langle 145,25,3 \rangle, $ \\ $ \langle 1,25,35 \rangle,\langle 15,24,345 \rangle, \langle 15,23,4 \rangle,\langle 12,4,5 \rangle, $ \\ $\langle 24,3,5 \rangle,\langle 14,34,5 \rangle $.
\end{itemize}
\subsubsection{Opposite Regulus Code from OR1}
We first consider the code obtained from OR1 by looking at the combinations of the spreads and opposite spreads.\\
$\bullet$ The regulus pattern of $mS1'^{o}$ is: $r_1 = \{l_1, l_5, l_9\}$, $r_2 = \{l_2, l_4, l_9\} $, $r_3 = \{l_3, l_6, l_9\}$ and $r_4 = \{l_7, l_8, l_9\}$. Evidently $mS1'^{o}$ is of type X with $l_9 = \{134,2,5u \}$ as the common line. So we have to find a permutation of the basis set $\{1,2,3,4,5\}$ which fixes $l_9 = \{134,2,5u \}$, cyclically permutes six of the remaining $8$ lines, and transposes the last pair. The following permutation, described as: $(12345)\rightarrow (45,134,234,145,u)$, acts on the lines as: $(l_1 \,l_6\,  l_7\, l_5\, l_3\, l_8)\,(l_2\, l_4)\,(l_9)$.\\
Hence the lines $l_1 = \{35,45,34 \} $, $l_3 = \{14,245,125 \}$, $l_5 = \{135,3u,234 \}$, $l_6 = \{4u,23,15 \}$, $l_7 = \{12,3,123 \}$, $l_8 = \{1,24,124 \}$ form the cyclic $S_6$, while $l_2 = \{13,345,145 \}$ and $l_4 = \{235,4,1u \}$ are the two lines making up the fourth regulus with $l_9$.\\
To construct the opposite Mps one replaces the lines of the fourth regulus $\{l_2, l_4, l_9\}$ with $l_{2}^{o} = \{13, 4, 134 \}$, $l_{4}^{o} = \{345, 1u, 2\}$ and $l_{9}^{o} = \{145, 235, 5u\}$. The regulus structure of the resulting Mps, denoted $E1$, is : $r_1 = \{l_1, l_5, l_{9}^{o}\}$, $r_2 = \{l_{2}^{o}, l_7, l_8 \}$, $r_3 = \{l_3, l_{4}^{o}, l_6\}$ and $r_4 = \{l_{2}^{o}, l_{4}^{o},l_{9}^{o}\}$. Evidently $E1$ is of type E, with each of the three lines of $r_4$ shared with exactly one of the other three reguli.\\
The code $E1 \cup (S2^{o})^{\perp}$ is sub-optimal with a single conflict: the line $l_4(E1) = \{345, 1u, 2\}$ is a subspace of the complement of $l_1(S2^{o})$, given by the span $\langle 1,2, 345 \rangle$. Applying the minimal change strategy, it is seen that no change in the $2$-vector of either subspace, keeping the Schubert cell structure, results in an optimal code. The allowed changes in the $3$-vector of the $2$-subspace fail as well. However the substitution of the vector $34 = 00110$ as the $3$-vector of the $3$-subspace yields another optimal code, denoted OR2. The new spread $mS2^{o}$ (modified $S2^{o}$) has regulus pattern: $r_1 = \{l_1, l_5, l_9\}$, $r_2 = \{l_1, l_6, l_8\} $, $r_3 = \{l_7, l_8, l_9\}$ and $r_4 = \{l_2, l_3, l_4\}$, where the lines are as below. Evidently it is of type I$\Delta$ with the fourth regulus as the distinguished regulus.\\
\emph{The Code OR2 and its Dual}
\begin{itemize}
\item $E1$: $ \{35,45,34 \}, \{13,4,134 \}, \{14,245,125 \},$\\$ \{345,1u,2 \}, \{135,3u,234 \}, \{4u,23,15 \}, \{12,3,123 \},$\\$ \{24,1,124 \}, \{145,235,5u \}$.\\
$(mS2^{o})^{\perp}$: $\langle 1,2,34 \rangle, \langle 145,25,35 \rangle,\langle 135,25,4 \rangle, $ \\ $\langle 15,25,45 \rangle,\langle 13,235,45 \rangle, \langle 125,3,4 \rangle, \langle 1,3,5 \rangle, $\\$\langle 24,34,5 \rangle,\langle 14,234,5 \rangle $.
\item The dual code:\\
$mS2^{o}$: $\{35,5,345 \}, \{14,4u,1u \}, \{13,235,125 \}, $ \\$ \{3, u, 3u\}, \{245,2u,123 \}, \{15,25,12 \},  \{2,4,24 \},$\\$ \{1,234,5u \}, \{134,23,124 \}.$ \\
$(E1)^{\perp}$: $\langle 1,2,345 \rangle, \langle 13,2,5 \rangle,\langle 145,25,3 \rangle, \langle 1,35,45 \rangle,$ \\ $\langle 15,24,345 \rangle, \langle 15,23,4 \rangle,\langle 12,4,5 \rangle, \langle 24,3,5 \rangle,$\\$\langle 14,245,345 \rangle $.
\end{itemize}
\emph{Remark:} The optimal code pairs reported above result from the application of the minimal change strategy on nearly optimal codes having the structure of the \dc. Another pair of optimal $(5,3)_2$ codes is the code $S1 \cup (S1')^{\perp}$ and its dual. This is a curious find, as the second spread $S1'$ is obtained from $S1$ as the result of modifying the nearly optimal EV code: $S1 \cup (S2)^{\perp}$. \emph{Hence we have a trio of Mps's: $S1$, $S1'$ and $S2$, such that two out of three possible pairs produce optimal codes by the \dc, while the pair $\{ S1,S1' \}$ does not}.
\section{Comparison with the Shortened Codes of Honold-Kiermaier-Kurz}
In this section we first analyze the construction in \cite{HKK}, described in Section \ref{sec:bg}, in terms of the spread types involved. The type of spreads is the basis for comparing our examples with the codes obtained by the HKK construction.\\
When restricted to $\ag$, the HKK construction yields optimal $(5,3)_2$ codes consisting of $9$ lines and an equal number of planes. This is evidently an instance of the {\dc}: the set of $9$ lines in $\ag$ ($2$-subspaces in $\bF_{2}^{5}$) forms an Mps and the set of $9$ planes ($3$-subspaces in $\bF_{2}^{5}$) is the dual of another Mps of lines. We now identify the types of the Mps's used in the line spread and the dual.
\subsection{The Type of the Line Spread}
The maximal partial spread $S$ obtained as the uncomplemented spread in the HKK construction is characterized over $\bF_q$ in \cite{HKK} as follows: the $q^k$ uncovered points (holes) of $S$ form the set-wise complement of a $k$-subspace $X_0 \in S$ in a $(k+1)$-subspace $Y_0$. The subspace $X_0$ is termed the \emph{moving subspace} of $S$, as it can be replaced by any $k$-subspace of $Y_0$ to form a maximal partial spread. To identify the type of Mps obtained as the line spread in the construction, when restricted to $\bF_2$, we state the following theorem of \cite{RS} (Theorem 3.2).
\bthm \label{thm:Xcls}
(i) The partial spread $S_8 = \{ \lambda_1, \lambda_2,  \cdots, \lambda_8 \}$ which arises from any partition of $PG(4,2)$ of the form $\lambda_1\cup \lambda_2 \cup \cdots\cup  \lambda_8 \cup \alpha_9  $, $\alpha_9$ a plane, is regulus-free.\\
(ii) If, in the above partition, $\lambda_9 $ is any line of the plane $\alpha_9$, then $S_9 = \{ \lambda_1, \lambda_2,  \cdots, \lambda_8, \lambda_9 \} $ is a partial spread in $PG(4,2)$ of type X. $\blacksquare$
\ethm
Based on the description in \cite{HKK} and the above theorem, we have the following
\bprop 
The line spread in the HKK construction is of type X.
\eprop
\begin{IEEEproof} From the description in the HKK construction, the ``moving line " $X_0$ is chosen from the plane $Y_0$, identified with $E_1 = (\Sc + P)\cap H$ in \cite{HKK}, which contains the holes of the spread (cf. Sections $3.2$ and $3.3$, \cite{HKK}). Therefore we have: $Y_0 = X_0 \cup \{\textrm{holes of} \,\, S_9\}$. The partial spread $S_8$ of $8$ lines is obtained by shortening those codewords of $\Gc$, the lifted Gabidulin code, which contain the point $P$. Hence the shortened lines forming $S_8$ are all disjoint from $Y_0$. So the partition $S_8 \cup Y_0 = PG(4,2)$ is of the type given in Theorem \ref{thm:Xcls} (i), and the line $X_0 \subset Y_0$ completes the Mps $S_9$ when added to $S_8$. Hence, the $S_8$ of lines is regulus-free, and the resulting $S_9$ is of type X.
\end{IEEEproof}
%An alternative argument, leading to the same conclusion, may be based on the discussion in Section $2.1$, pg. $5$, of \cite{HKK14}. There it is argued that if a line $M$ belonging to an Mps of size $9$ lies in the plane containing the holes of the Mps, $M$ belongs to all $4$ reguli of the Mps. This also implies that all the other lines in that Mps, which are disjoint from $M$, belong to exactly one regulus each. Therefore, the Mps in question is of type X as described in \cite{RS} (referred in \cite{HKK14}). Clearly $X_0 \subset Y_0$ fits the above description.
%Moreover, for $q=2,k=2$,(cf. Sections 3.2 and 3.3 of \cite{HKK}) $8$ lines of the spread $S$ form a partition of $\ag$ with the plane $Y_0$ containing the ninth line $X_0$.\\
\subsection{The Type of the Dual Spread}
For identifying the type of the dual spread, we refer to Remark $4$ of Section $3.3$ in \cite{HKK}. There it is asserted that the spread obtained in the dual form has the same type as the uncomplemented spread. We give a proof of this assertion by connecting the construction of the dual spread with the classification results in \cite{RS}.\\
Recall that the dual spread of planes obtained in the HKK construction combines $8$ codewords of the lifted $(6,3,2)$ Gabidulin code $\Gc$ with another additional plane, denoted $E' = E_2 $ in \cite{HKK}. The set of $8$ codewords are disjoint from a special plane $\Sc$ in $PG(5,2)$ which meets the shortening hyperplane $H = \ag $ in a line. The ninth plane $B_9 = E_2$ in the dual spread also meets the plane $\Sc$ in a line, denoted by $L_2$ in \cite{HKK}. We begin by identifying $B_9$ as the dual of the moving line in a type X spread and establish that the remaining planes $ B_i, \, i = 1, \cdots, 8 $, can be identified as the duals of lines $ \lambda_i, \, i = 1, \cdots, 8 $, of the spread, in some order.
\bprop
The spread obtained in dual form in the HKK construction of optimal $(5,3)_2$ codes is of type X.
\eprop
\begin{IEEEproof}
Recall that a maximal partial line spread of type X arises out of a partition of $PG(4,2)$ of the form \cite{RS}: 
\beq \label{eq:xsprd} 
S= \lambda_1\cup \lambda_2 \cup \cdots\cup  \lambda_8 \cup \alpha_9 
\eeq
where $\alpha_9$ is a plane. The ninth line of the spread is any line $\lambda_9 \subset \alpha_9$ and so, the $4$ holes of the spread are given by $\{h_1, h_2, h_3, h_4\} = \alpha_9 \setminus \lambda_9$. The duals of the holes $h_i, \, i = 1,\cdots , 4$, are hyperplanes in $PG(4,2)$, i.e. solids denoted by $S_i,  \, i = 1,\cdots , 4$. Therefore, we have:
\beqn
\left( \bigcup_{i=1}^{4} h_i \right) ^{\perp} =  \bigcap_{i=1}^{4} {\left( h_i \right) ^{\perp}} = \bigcap_{i=1}^{4} S_i
\eeqn
If the ninth plane $B_9$ is identified as ${\lambda_9}^{\perp}$, where $\lambda_9$ is as described above, it follows that $L_2 \subset B_9 $ can be identified as the meet of the dual solids: \beqn
L_2 = \left( \bigcup_{i=1}^{4} h_i \right) ^{\perp} = \bigcap_{i=1}^{4} S_i
\eeqn
Remains to identify the planes $B_i $ as $ {\lambda_i}^{\perp}, \, i = 1, \cdots, 8$. By duality, it is enough to prove that:
\beq
\dim (B_i \cap L_2) = 5 - \dim \left(\lambda_i \cup \left( \bigcup_{i=1}^{4} h_i \right) \right); \, i= 1, \cdots, 8.
\eeq
We have $\dim \left(\lambda_i \cup \left( \bigcup_{i=1}^{4} h_i \right) \right) = 5$ for all $i$. This uses the fact that $h_i, i = 1, \cdots, 4$, are the holes of a maximal partial spread, forming the complement of $\lambda_9$ in the plane $\alpha_9$ in Eqn (\ref{eq:xsprd}). Hence any three of them are linearly independent and the union of the holes with any $\lambda_i,\, i = 1,\cdots, 8$, disjoint from $\alpha_9$, spans the entire ambient space.\\
But by the HKK construction, $B_i,\, i = 1,\cdots,8$, are all disjoint from $L_2 \subset \Sc$, the special plane disjoint from all the codewords of $\Gc$. So $\dim (B_i \cap L_2) =0$ for all $i$, and the assertion is proved.\\
Hence the set of planes obtained by the HKK construction of $(5,3)_2$ codes is the dual of a line spread of type X. 
\end{IEEEproof}
It follows that both the line spreads obtained by the HKK construction for optimal $(5,3)_2$ codes are of type X. The spread types of our reported examples are as follows:
\ben 
\item The MEV code and its dual are constituted of two Mps's both of type I$\Delta$; 
\item The code (and its dual) formed of the EV code line spread and the MEV line spread use Mps's of types X and I$\Delta$, respectively; 
\item The code OR1 and its dual use Mps's of types X and I$\Delta$, respectively;
\item The code OR2 and its dual use Mps's of types E and I$\Delta$, respectively. 
\een
\emph{Therefore, based on the types of spreads used, all our examples fall outside the framework of the HKK construction.}
\section{Conclusion and Future Work}
We have presented several optimal binary $(5,3)$ projective space codes, which have been mostly obtained by the application of the strategy of minimal changes on nearly optimal codes. These nearly optimal codes were produced by replacing certain special reguli in the constituent spreads with the opposite reguli. The limitations of the minimal change strategy is obvious - it can take care of only a single conflict among the lines and planes in a nearly optimal code. But the fact that it has produced optimal codes with a vastly reduced search begs the following question: can the study of Schubert cell composition of spreads used in optimal codes lead to a general construction?\\
The paper of Honold \emph{et al.} \cite{HKK} represents a significant milestone in the search for optimal projective space codes. Their bound-achieving construction produces, in the binary case, codes involving just one of the three possible spread types. We have shown that our examples lie outside their framework in this respect. It will be worthwhile to modify or extend the HKK construction to encompass other spread types as well.\\
In \cite{HKK}, instances of the {\dc} yielding optimal $(7,34,5)_2$ codes have been reported. Another direction of future research would be to use the study of partial spread structures in the higher dimensions as well as for non-binary cases to obtain $(2k+1,2k-1)$ optimal codes in $PG(2k,q)$.    
\section*{Acknowledgment}
The author gratefully acknowledges the valuable discussions with Smarajit Das and Sumanta Mukherjee.

\end{document}